# The Boltzmann fair division for distributive justice

**Ji-Won Park[1,2]*, Jaeup U. Kim[3], Cheol-Min Ghim[3], Chae Un Kim[3]***

[1]Department of Regional Science, Cornell University, Ithaca, NY 14853, USA

[2]Department of Economics, University of Ulsan, Ulsan 44610, Korea

[3]Department of Physics, Ulsan National Institute of Science and Technology (UNIST), Ulsan 44919, Korea

* **Corresponding authors: Ji-Won Park and Chae Un Kim**.

Email: jp429@cornell.edu

Email: cukim@unist.ac.kr

**Abstract**

Fair division is a significant, long-standing problem and is closely related to social and economic justice. The conventional division methods such as cut-and-choose are hardly applicable to real-world problems because of their complexity and unrealistic assumptions about human behaviors. Here we propose a fair division method from a completely different perspective, using the Boltzmann distribution. The Boltzmann distribution adopted from the physical sciences gives the most probable and unbiased distribution derived from a goods-centric, rather than a player-centric, division process. The mathematical model of the Boltzmann fair division was developed for both homogeneous and heterogeneous division problems, and the players' key factors (contributions, needs, and preferences) could be successfully integrated. We show that the Boltzmann fair division is a well-balanced division method maximizing the players' total utility, and it could be easily fine-tuned and applicable to complex real-world problems such as income/wealth redistribution or international negotiations on fighting climate change.

Achieving social and economic justice for its members is among the most cherished goals of a civilized society [1]. Fair division of society's benefits (or burdens) is a long-standing, significant problem and is closely related to social and economic justice [2-22]. Fair-division problems seek to divide a valuable resource between multiple players based on a certain notion of fairness. Such problems have many practical applications, including international border settlements, divorce and inheritance settlements, allocation of shared computational resources [23], and distribution of scarce medical resources (e.g., vaccines) with respect to pandemics, income and wealth redistribution, and tax policy. The fair-division problem is sometimes called the cake-cutting problem [24], in which a single divisible good (represented by a cake) should be divided between multiple players fairly and efficiently.

The cake-cutting problem has attracted the attention of mathematicians, computer scientists, economists, and political scientists since the 1940s [9,12,13,21,25-31]. Various notions of fairness have been proposed to solve such problems [7,15,19]; of these, the most widely used are envy-freeness and proportionality [31]. Under envy-free division, no player should envy other players for their cake allocations. Under proportional division, each player should receive cake at least in proportion to certain criteria, such as $1/n$ ($n$ is number of players), contributions to the production of cake, or needs for the cake.

Since studies of the cake-cutting problem in the 1940s due to Steinhaus and Knaster [7,21,32], the best-known method for cake-cutting is cut-and-choose, based on envy-freeness [33,34]. Steinhaus observed that the cut-and-choose protocol could be extended to three players and asked whether it could be generalized to any number of players [12,35]. More recently, Aziz and Mackenzie solved the cake-cutting problem by proposing a discrete and bounded envy-free protocol for any number of players and showing that the maximum number of queries required by the protocol is $n^{n^{n^{n^{n^{n}}}}}$ [36].

In general, most existing algorithms for cake-cutting become drastically complex when three or more players participate [37]. Further, in reality, it should be considered that the cake could be a heterogeneous good, and the participating players may have personal preferences for different parts of it. The heterogeneity of the cake and players' personal preferences further complicate cake-cutting problems [7].

Beyond the complexities, existing cut-and-choose procedures have a fundamental limitation. Previous algorithms for these problems seek a solution in which all players believe they have received a fair share so that they do not envy the other players. These algorithms impose several assumptions on the participating players [38]. For example, they implicitly assume that all players act rationally, without emotional decisions or irrational behaviors, and do not try to manipulate the game. They also assume that players lack prior information about the other players' preferences and choices. However, these assumptions are often impractical and unrealistic, as there always could be greedy, selfish, impulsive, irrational, and cheating players participating in the procedures [39,40]. As a result, cut-and-choose fair division cannot easily lead to finding peaceful negotiations in today's major social or economic problems such as income/wealth redistribution.

In contrast to the protocols based on envy-freeness, fair division based on proportionality is rather simple [21] and is often used as a common criterion for distributive justice. The fairness notion of proportionality often assumes that all players have equal rights to the cake, and thus everyone agrees that everyone else is entitled to a fair share, that is, at least $1/n$. Within this assumption, envy-freeness and proportionality could be equivalent when there are only two players. However, when more than three players are involved, envy-free divisions could be proportional, but not vice versa [7]. On the other hand, Aristotle identified distributive justice with "proportionate equality": that is, the view that goods should be distributed not strictly equally but in proportion to their worth or merit [41]. In the same vein as Aristotle, weighted proportionality was recently proposed in which the participating players have different rights to the cake [42]. Fair division based on proportionality is closely related to the key principles of distributive justice: justice as equality (egalitarianism), justice based on contribution (capitalist justice), and justice based on needs (socialism) [43].

Fair division based on proportionality has also had its critics. According to the proportionality on equality (egalitarian), goods should be allocated to people in equal portions. Critics argue that egalitarianism ignores some characteristics that should be taken into account in distributing goods such as contributions and needs, therefore leading to a decline in society's productivity and efficiency [43,44]. On the other hand, critics of capitalist proportionality claim that some people may be ill, handicapped, too old or too young, or otherwise incapable of contributing anything through personal efforts, and that their minimal needs for the goods cannot be met, therefore leading to the decline in the overall society's welfare [43]. Finally, critics of socialist proportionality point out that there would be no relationship between the amount of contribution a person makes and the amount of rewards they receive. They argue that fair division based on socialist proportionality is against human nature, which is essentially self-interested and competitive, therefore leading to declining productivity [43,45,46].

In order to overcome the complexities and limitations of the existing methods, a fair division from a completely different perspective is required. In this article, we propose a fair division method inspired by the physical principle, the Boltzmann distribution. The mathematical model of the Boltzmann fair division was developed for both homogeneous and heterogeneous cake-cutting problems, where players' contributions, needs, and preferences for heterogeneous portions could be successfully integrated. Then the Boltzmann fair division was optimized by maximizing the players' total utility. Through the analysis of empirical data, we found that the Boltzmann fair division is a well-balanced division method beyond the conventional division methods.

The Boltzmann fair division is simple but highly versatile, and can be easily fine-tuned to apply to a variety of division problems. Thus, we believe that our work makes a seminal contribution to the feasible fair division for complex, real-world problems.

## Results

Conventional approaches to the cake-cutting problem seek a solution in which all participating players believe they have received a fair share. In reality, however, it is almost impossible to find such a solution, especially when players contribute different amounts to producing the cake, and different players have different amounts of need for it. If finding a fair division that makes all players "feel" satisfied is practically impossible to achieve, we need to search for a fair solution from a completely different perspective. Below, we propose a fair division in which players' subjective feelings are completely excluded. Instead, we allow the divisible goods to be divided spontaneously among participating players naturally and unbiasedly. This unbiased (or fair) cake division can be achieved by using the Boltzmann distribution [47,48] .

In the physical sciences, the Boltzmann distribution describes the equilibrium probability distribution of physical particles in a physical system's energy substates [49]. The description is valid in a classical physics regime in which each particle of the system is identical to but distinguishable from the others, and the interactions between them are negligible. In the Boltzmann distribution, the probability $P_j$ that a particle can be found in the $j^{th}$ substate is inversely proportional to the exponential function of the substate energy $E_j$ (*i.e.*, $P_j \propto e^{-\beta E_j}$ , $\beta = 1/kT$ ($k$: Boltzmann constant, $T$: absolute temperature)). The Boltzmann distribution is based on entropy maximization and provides the most probable, natural, and unbiased distribution of a physical system at thermal equilibrium.

In this paper, we apply the Boltzmann distribution to the cake-cutting problem. The concept of the physical particle is replaced by the cake unit (infinitesimal piece of the cake). The idea of the physical substates is replaced by the participating players. Then, the probability that a cake unit is allocated to a player *j* is proportional to the exponential function of the division potential $E_j$, *i.e.*, $P_j \propto e^{\beta E_j}$, where $\beta$ is a positive constant. The division potential $E_j$ of player *j* could be a measure of the player's talent or contribution to the cake's production. Note that the probability exponent has a positive sign in front of $\beta E_j$. Based on this definition, when a player has a higher division potential, cake units would be more likely to be distributed to the player. Details on the Boltzmann fair division developed for both homogeneous and heterogeneous cake-cutting problems can be found in Methods.

To demonstrate optimal cake-cutting using the Boltzmann fair distribution, we performed empirical analysis on five players. We assumed that the five players make different contributions (5%, 10%, 20%, 25%, and 40%; total sum is 100%) to produce the cake, and have different needs (4%, 10%, 24%, 34%, 53%; total sum is 125%; Table S1). The total number of cake units $N_j$ allocated to a player *j* was normalized to the allocated share of cake as a percentage.

In the case of homogeneous cake-cutting (Fig. 1), cake was allocated to players based on the Boltzmann distribution $P_j$ (Eq. 1 in Methods), where each player's contribution was used as the division potential $E_j$. Then, the optimal division constant $\beta^*$ was determined at the maximum point of the total utility function (Eq. 9 in Methods). Hyperbolic tangent function was used as the players' utility function (Eq. 8 in Methods), with the player's need for the cake as the need values $D_j$ so that the following condition was satisfied: $u_j(0) = 0$, $u_j(D_j) = \tanh(1) \cong 0.762$ (Fig. S1). This means that if a player receives what they need ($D_j$), their utility is ~76.2% saturated. Detailed information about the numerical calculations is in Table S2.

Fig. 2 shows the total utility of five players as a function of $\beta$. The total utility $U(\beta)$ increases rapidly to reach an optimal value as $\beta$ increases up to $\beta^* = 0.029$, and then gradually decreases. Interestingly, the total utility at $\beta = 0$ is higher than when the $\beta$ value is large, and this result shows that the completely uniform cake division (i.e., egalitarian) has a higher total utility than the significantly biased cake division.

The optimal cake division using the Boltzmann distribution is summarized in Fig. 3 and Table 1. The results were compared with the three conventional division criteria, Egalitarian (uniform cake division), Proportionality I (contribution), and Proportionality II (need). As shown in Table 1, Player 3, having the middle contribution and need, receives similar amounts of cake (~20%), independent of the four criteria: Boltzmann, Egalitarian, Proportionality I (contribution), and Proportionality II (need). By contrast, other players receive different amounts of cake according to these four criteria.

Players 1 and 2, having relatively low contributions and needs, receive the largest amount of cake under Egalitarian. On the other hand, Players 1 and 2 receive the smallest amount of cake under Proportionality II (need) because of their small needs compared with the other players, leading to the largest cake deficiency (received share – need) among four criteria. The preferred criteria for Players 1 and 2 are, in order, Egalitarian, Boltzmann, Proportionality I (contribution), and Proportionality II (need). In contrast, Players 4 and 5, having relatively high contributions and needs, receive the largest amount of cake under Proportionality II (need), resulting in the lowest deficiency. On the other hand, Players 4 and 5 receive an amount of cake equal to that of other players under Egalitarian, leading to the highest deficiency. The preferred criteria for Players 4 and 5 are, in order, Proportionality II (need), Proportionality I (contribution), Boltzmann, and

Egalitarian.

The results show that the Boltzmann distribution is the criterion neither most nor least preferred by the participating players. Yet, of the four criteria, it brings the highest total utility. Note that the range of divided cake shares between five players is smaller in the Boltzmann than in the two Proportionality criteria, which indicates that the Boltzmann distribution is a *balanced division* between the Egalitarian and the Proportionality criteria (Fig. 3).

In the case of heterogeneous cake-cutting, the cake has equal portions of four flavors, vanilla, chocolate, strawberry, and broccoli (Fig. 1). The five players' preferences for the four flavors are given with the weight factor values (Table S1). The cake was allocated to players based on the weighted Boltzmann distribution probability $P_j^i$ (Eq. 5 in Methods). As in the homogenous cake-cutting, each player's contribution and need were used as the division potential $E_j$ and the need value $D_j$. Then, the optimal division constant $\beta^*$ in the Boltzmann distribution was determined at the maximum point of the total utility function (Eq. 9 in Methods). Detailed information about the numerical calculations is in Table S3. Fig. 2 shows the total utility of the five players as a function of $\beta$. As in the homogeneous case, total utility rapidly increases to an optimal $\beta$ value of 0.029, and then gradually decreases.

Heterogeneous cake division using the weighted Boltzmann distribution is summarized in Fig. 4 and Table 2. Compared with homogeneous cake-cutting, the results of heterogeneous cake-cutting are affected not only by the players' contributions/needs but also by their flavor preferences. The major difference in heterogeneous cake-cutting is manifested in the comparison between Player 3 and Players 1 and 4. Player 3 has a strong preference for chocolate; therefore, the player's weight value for chocolate is set at 1, and the values for the other flavors are 0 (Table S1). Chocolate is also preferred by three other players. Thus, although the largest share of the chocolate-flavored cake is allocated to Player 3, the total allocated share of cake is significantly smaller than in the homogeneous cake-cutting. On the other hand, Players 1 and 4 prefer the least popular flavor, broccoli (Table S1); therefore, all the broccoli-flavored cake was allocated to them. Thus, Players 1 and 4 receive a larger share of cake than in the homogeneous cake-cutting, and Player 1 receives even a larger share of cake than Player 2. Overall, the results show that the weighted Boltzmann distribution can smoothly handle complicated cake-cutting situations by incorporating a variety of factors, such as contributions, needs, and preferences for heterogeneous portions.

**Discussion**

Distributive justice is concerned with the fair division of society's benefits (or burdens) among people [50]. Questions of distributive justice arise when different people make claims on society's benefits and not all claims can be satisfied. The central cases are those where there is a scarcity of benefits compared with the numbers and desires of those who want these goods. The final goal of fair division is to make all players believe they have received a fair share. However, what each player thinks is a fair share could be subjective and might differ depending on whom each player shares it with. The degree of fair share that each player perceives might differ depending on the degree of players' factors such as talents, contributions, needs, or personal preferences. For example, when a cake is given to a group of players and none of the players have contributed to its production, it could be natural to assume that everyone is equally entitled to it. However, when the players have made different contributions to the cake's production, an asymmetry is inherent in the problem, and it becomes drastically difficult to reach a harmonious solution. The most critical issues of distributive justice are concerned with how to divide social goods when participating players have different factors (such as contributions or needs) with respect to the goods and display egocentric and irrational behaviors.

In this paper, we proposed a method for solving the cake-cutting problem from a different perspective, using the Boltzmann distribution. The Boltzmann distribution is based on the entropy maximization that nature follows at its thermal equilibrium. Note that in a fair division using the Boltzmann distribution, participating players cannot subjectively claim how much and which parts they should receive. Instead, the cake-units are spontaneously distributed to players purely depending on the Boltzmann probability. This spontaneous distribution is the modus operandi of the maximum entropy principle that underlies the most equitable and impartial distribution of resources in nature. Therefore, in the Boltzmann fair division, players' emotional, irrational decisions or cheating behaviors are fundamentally limited.

We believe that fair division using the Boltzmann distribution has several superior aspects compared with conventional methods. Compared with the fair division based on envy-freeness (such as cut-and-choose), the Boltzmann fair division can easily apply to otherwise extremely complicated cases, where heterogeneous goods should be fairly divided among multiple players who make different contributions to their production and different preferences for different parts of the goods. Compared with the fair divisions based on proportionality (Egalitarian, Capitalist, and Socialist), the Boltzmann fair division can simultaneously integrate various key factors of the participating players, such as contributions or needs.

Note that the Boltzmann fair division involves total utility maximization; therefore, it pursues the maximization of social welfare. Our empirical analysis shows that it is a balanced solution compared with the conventional fair divisions. It should be emphasized that the Boltzmann probability takes the form of the exponential function $P_j \propto e^{\beta E_j}$. Thus, even when a player's contribution is 0, the Boltzmann probability becomes a non-zero value. As opposed to the Boltzmann fair division, in the Capitalist proportionality, no goods would be allocated to the player.

The result shows that the Boltzmann fair division is favorable to the socially disadvantaged or underprivileged; therefore, it could serve as a guideline to reduce inequality in various social and economic problems.

The Boltzmann fair division in this study can be further fine-tuned for application to real-world division problems. The hyperbolic tangent utility function in our study could be modified with other forms of functions to reflect a more realistic utility of the players. For example, the utility function can be further generalized by allowing its maximum saturation value to vary (Fig. S2). Although it is not very sensible to differentiate an individual's utility maximum, it may well find justification for a group of individuals or communities having distinct population sizes and/or value systems. Another meaningful extension of the utility function is to introduce sigmoidal behavior with an inflection point at $x > 0$ to capture the transitional characteristics of the marginal utility. In fact, most of the named cumulative probability distributions, including normal, gamma, and logistic functions, are sigmoidal at least for certain combinations of their scale and shape parameters. Additional considerations of the Boltzmann probability and total utility maximization are described in the Supplementary Information.

As proof of principle, we considered only five players in our empirical analysis, but it will be straightforward to extend our model to a larger number of players. The division potential ($E_j$) and need ($D_j$) in the Boltzmann distribution could be modified by considering not only contributions and needs but also factors such as players' talents, efforts, and physical properties (age and health conditions). It would also be possible to extend our model from the fair division of beneficial goods to the fair division of burdens. For this purpose, the contributions ($E_j$) and needs ($D_j$) in our study would be replaced by players' responsibilities and abilities. The fine-tuned Boltzmann fair divisions could then be applicable to a variety of social and national issues, such as negotiations on climate change between developed and developing nations.

In conclusion, we proposed a fair division using the Boltzmann distribution. We showed that it can be modeled over the complexities and limitations of conventional fair divisions and successfully applied to realistic division problems. We anticipate that our model could serve as a feasible fair division for complex real-world problems, leading our societies one step closer to distributive justice.

## Methods

Here we describe mathematical models of the Boltzmann fair division for the cake-cutting problems. First, we consider a simple case, in which the cake is a homogeneous good. Then, we extend our model to a more general case of heterogeneous goods. Finally, we optimize the Boltzmann division by maximizing the players' total utility.

### *Cake-Cutting Using the Boltzmann Distribution: Homogeneous Case*

Suppose that $n$ players are participating in the cake-cutting problem. Consider that the cake is homogeneous and hypothetically cut into $\tilde{N}$ cake units, where $\tilde{N}$ is a large number so that the cake unit is infinitesimally small. A player $j$ has cake-division potential $E_j$, which reflects the contribution of player $j$ to producing the cake. Based on the Boltzmann distribution, the probability $P_j$ that a cake unit is allocated to player $j$ is proportional to the exponential function of the division potential $E_j$ (i.e., $P_j \propto e^{\beta E_j}$, $\beta$ is a division constant ($\beta \geq 0$)). Then the total number of cake units that are divided into player $j$ is $N_j = \tilde{N} \times P_j$.

Considering the normalization condition ($\sum_{j=1}^{n} P_j = 1$), the probability $P_j$ and the total number of cake units $N_j$ allocated to player $j$ (= 1,2,…, $n$) can be expressed as

$$P_j = \frac{e^{\beta E_j}}{\sum_{k=1}^{n} e^{\beta E_k}} \tag{1}$$

$$N_j = \tilde{N} \times P_j = \frac{\tilde{N} e^{\beta E_j}}{\sum_{k=1}^{n} e^{\beta E_k}} \tag{2}$$

*Cake-Cutting Using the Boltzmann Distribution: Heterogeneous Case*

Now, we want to extend our model to a more complicated case, in which the divisible good is heterogeneous. Let us assume that the cake has $m$ different flavors such as vanilla, chocolate, strawberry, broccoli, and so forth. Each of $n$ participating players may have different preferences for these flavors, and the preference of player $j$ for flavor $i$ can be expressed with a weight factor, $w_j^i$ [51,52]. Then the probability $P_j^i$ that a cake unit of flavor $i$ is allocated to player $j$ depends not only on the contribution of player $j$ to producing the cake but also on the player's preference for flavor $i$.

$$P_j^i \propto w_j^i e^{\beta E_j} \tag{3}$$

$$\sum_{i=1}^{m} w_j^i = 1 \text{, for each player } j \tag{4}$$

With the constraint on $w_j^i$ (Eq. 4), the weighted Boltzmann distribution reduces to the Boltzmann distribution for the homogeneous case, when $w_j^i = 1/m$ for any $i$ and $j$.

Considering the normalization condition ($\sum_{j=1}^{n} P_j^i = 1$), the probability ($P_j^i$) can be expressed as follows.

$$P_j^i = \frac{w_j^i e^{\beta E_j}}{\sum_{k=1}^{n} w_k^i e^{\beta E_k}} \tag{5}$$

Then the total number ($N_j^i$) of cake units with flavor $i$ allocated to player $j$ is

$$N_j^i = \tilde{N}^i \times P_j^i \tag{6}$$

where $\tilde{N}^i$ is the total number of cake units with flavor $i$, s.t. $\sum_{i=1}^{m} \tilde{N}^i = \tilde{N}$.

Hence, the total number of cake units ($N_j$) allocated to player $j$ is

$$N_j = \sum_{i=1}^{m} N_j^i \text{, for } j = 1,2, \cdots, n \tag{7}$$

*Optimal Cake Division via Utility Maximization*

The cake division $\{N_1, N_2, \cdots, N_n\}$ obtained from the Boltzmann distribution is simple, with the single adjustable division constant $\beta$, yet highly versatile. If the division constant $\beta$ approaches 0, all players receive an equal amount of cake, representing uniform cake division (egalitarian). When $\beta$ increases to a large value, only a few players having made the highest cake contributions receive most of the cake, representing a highly biased cake division. Thus, cake-cutting using the Boltzmann distribution can represent a wide range of cake divisions, from an idealistic uniform division to a highly biased division.

Now we want to determine the $\beta$ value for the optimal cake division in which the total sum of each player's utility, $U = \sum_{j=1}^{n} u_j$, is maximized. The player's utility function $u_j$ should reflect the realistic welfare (well-being, happiness, and satisfaction) that each player feels as the number of allocated cake units increases. The utility function would start at 0, and increase rapidly as the number of allocated cake units increases. However, as the number of allocated cake units increases beyond the player's need, the utility would be saturated, after which it would increase rather slowly. This characteristic of individual utility can be modeled using various mathematical functions with an asymptotic saturation point. Of the numerous choices commonly made in biochemistry, ecology, the natural sciences, and economics, we chose, for its analytical simplicity, the hyperbolic tangent function with individual need as a scale parameter: that is,

$$u_j(x) = \tanh\left(\frac{x}{D_j}\right), \text{ for } j = 1,2,\cdots,n \tag{8}$$

where $x$ is the share of allocated cake units and $D_j$ is a constant reflecting player $j$'s need of the cake.

Therefore, we want to find $\beta^*$ which maximizes the total utility, that is,

$$\beta^* = \arg\max_{\beta} U(N_1, N_2, \ldots, N_n) = \arg\max_{\beta} \sum_{j=1}^{n} u_j(N_j) = \arg\max_{\beta} \sum_{j=1}^{n} \tanh\left(\frac{N_j(\beta)}{D_j}\right) \tag{9}$$

subject to

$$N_j(\beta) = \begin{cases} \tilde{N} \times \dfrac{e^{\beta E_j}}{\sum_{k=1}^{n} e^{\beta E_k}} & (\text{for homogeneous case}) \\ \sum_{i=1}^{m} N_j^i = \sum_{i=1}^{m} \left( \dfrac{\tilde{N}^i w_j^i e^{\beta E_j}}{\sum_{k=1}^{n} w_k^i e^{\beta E_k}} \right) & (\text{for heterogeneous case}) \end{cases} \tag{10}$$

In practice, the numerical value of $\beta^*$ can be found from the extremum condition:

$$\frac{\partial U}{\partial \beta} = \sum_{j=1}^{n} \frac{\partial U}{\partial N_j} \cdot \frac{\partial N_j}{\partial \beta} = 0 \quad (11)$$

Then, the obtained cake division $\{N_1, N_2, \cdots, N_n\}$ for $\beta^*$ becomes the optimal fair division dictated by the Boltzmann distribution.

**Acknowledgments**

This work was supported by the Samsung Science and Technology Foundation (SSTF-BA1702-04).

**Author Contributions**

J.-W.P. conceived the research, J.-W.P. and C.U.K. developed the theoretical models, ran the empirical analysis, analyzed the results. J.-W. P., J.U.K., C.-M.G. and C.U.K. contributed to the overall interpretation and wrote the paper.

**Competing Interests:** The authors declare no competing interests.

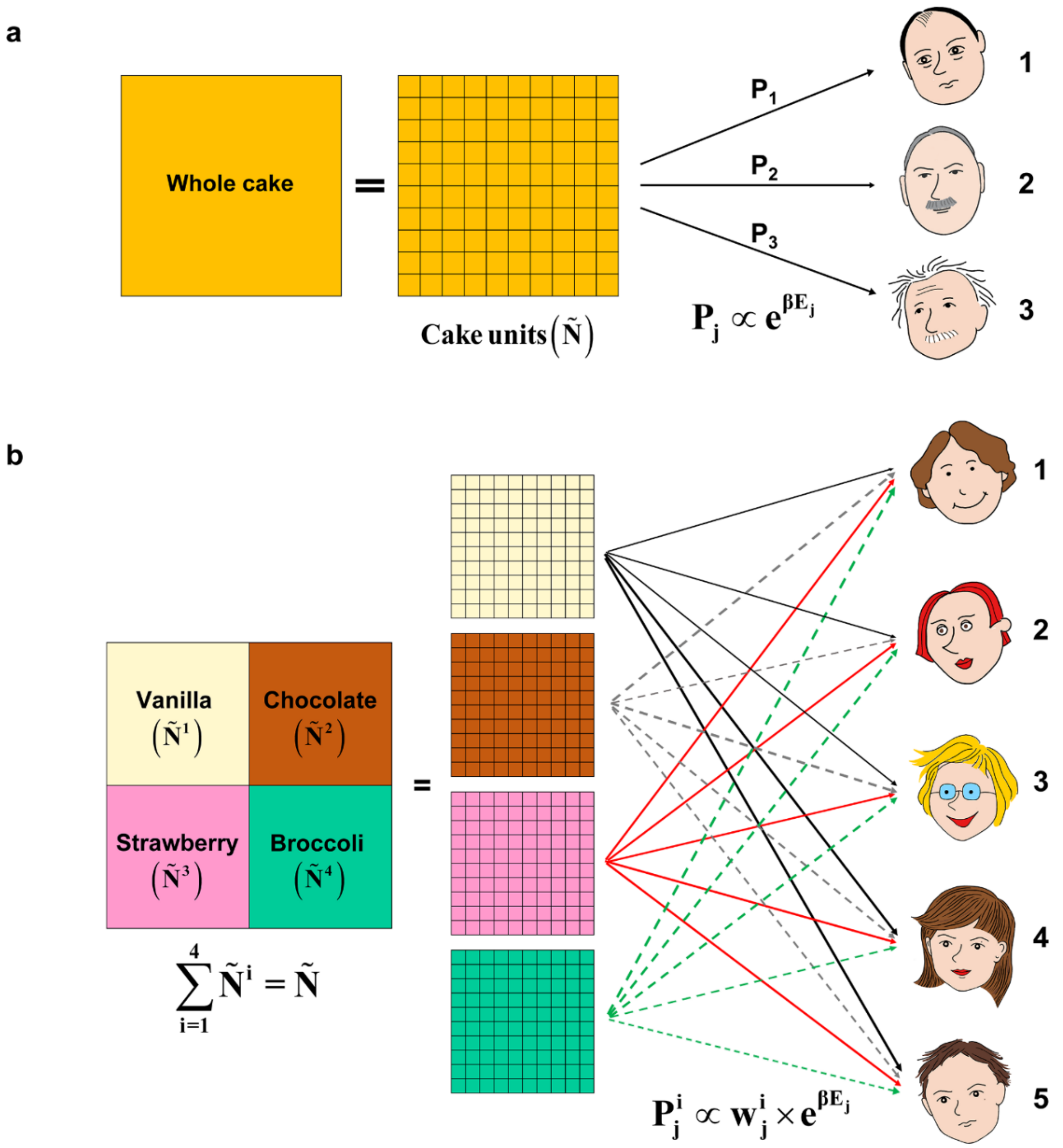

**Fig. 1 | Boltzmann fair division for homogeneous and heterogeneous cake-cutting problems.** **a** Homogeneous cake cutting, in which $\tilde{N}$ is the total number of cake units, $E_j$ is the division potential of player $j$, and $P_j$ is the Boltzmann probability that a cake unit is allocated to player $j$. **b** Heterogeneous cake cutting, in which $\tilde{N}^i$ is the total number of cake units with flavor $i$, $w_j^i$ is the weight factor expressing player $j$'s preference for flavor $i$, and $P_j^i$ is the Boltzmann probability that a cake unit of flavor $i$ is allocated to player $j$.

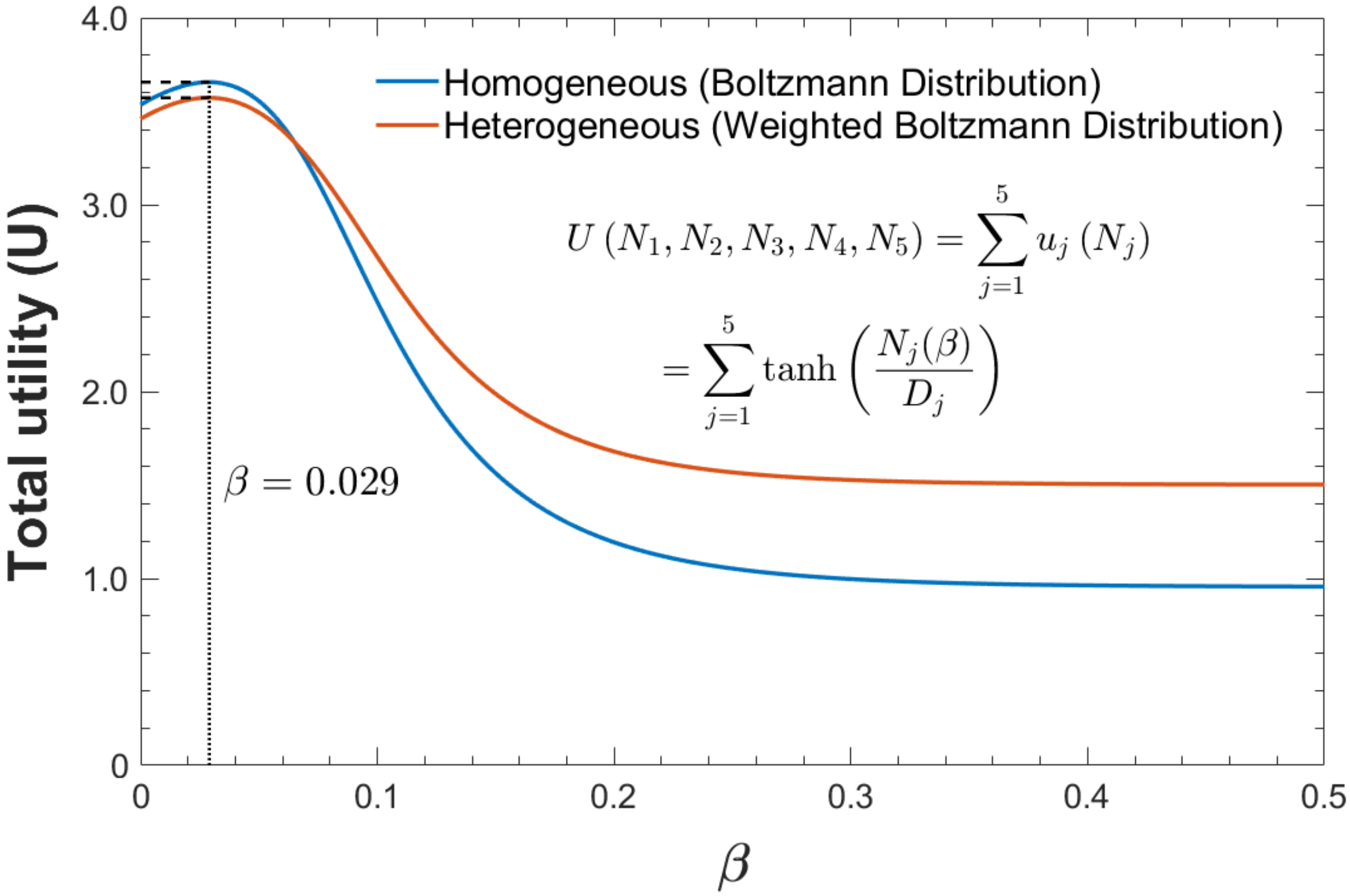

**Fig. 2 | Total utility (U) for the homogeneous and heterogeneous cake-cutting as a function of $\beta$.** The homogeneous (blue) and heterogeneous (red) cases show maximum at $\beta = 0.029$, and then the utilities gradually decrease.

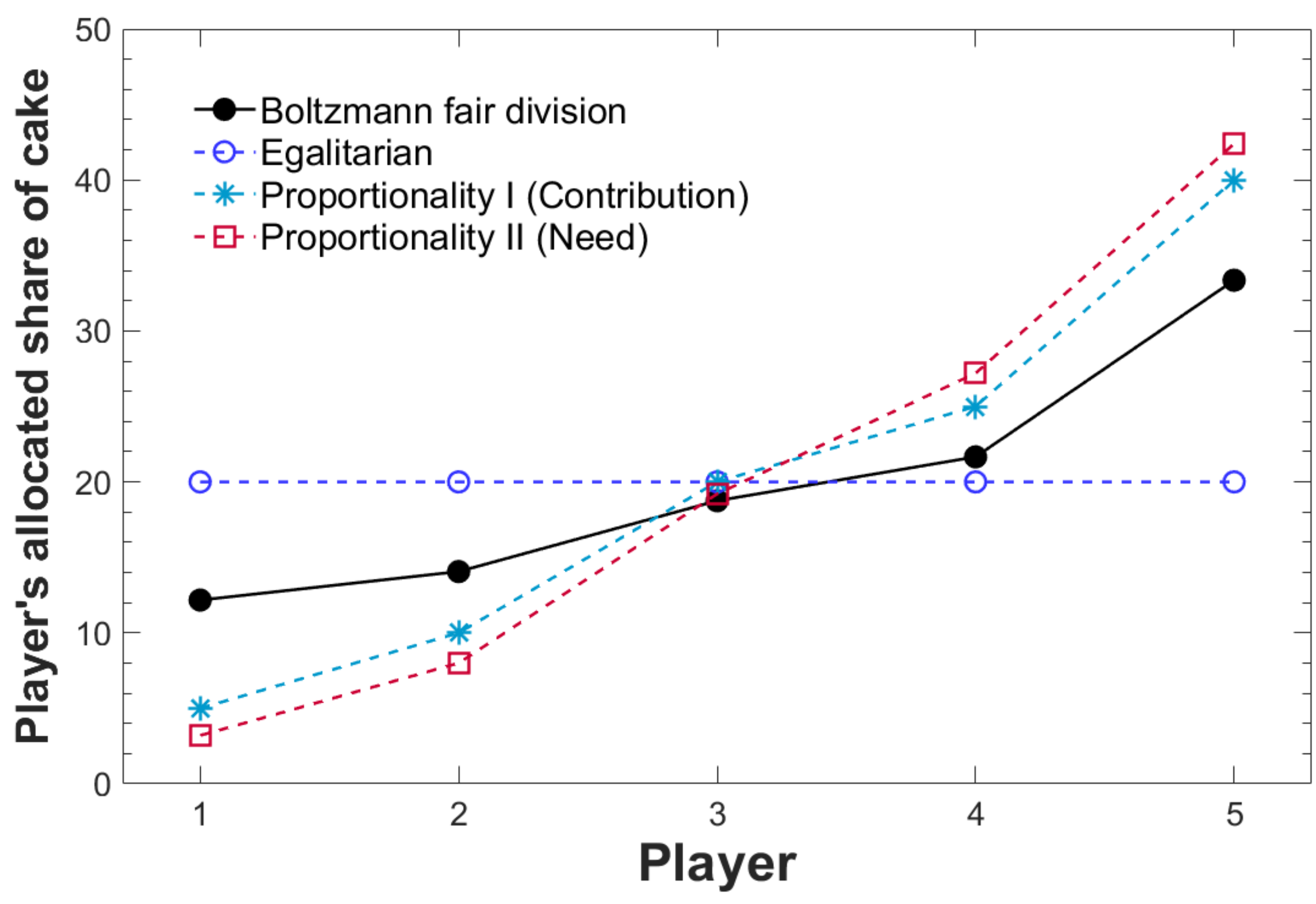

**Fig. 3 | Allocated share of cake under homogeneous cake-cutting.** The results using the Boltzmann are compared with three conventional division criteria, Egalitarian (uniform cake division), Proportionality I (contribution), and Proportionality II (need). Note that the Boltzmann fair division shows a well-balanced division between Egalitarian and Proportionality I and II.

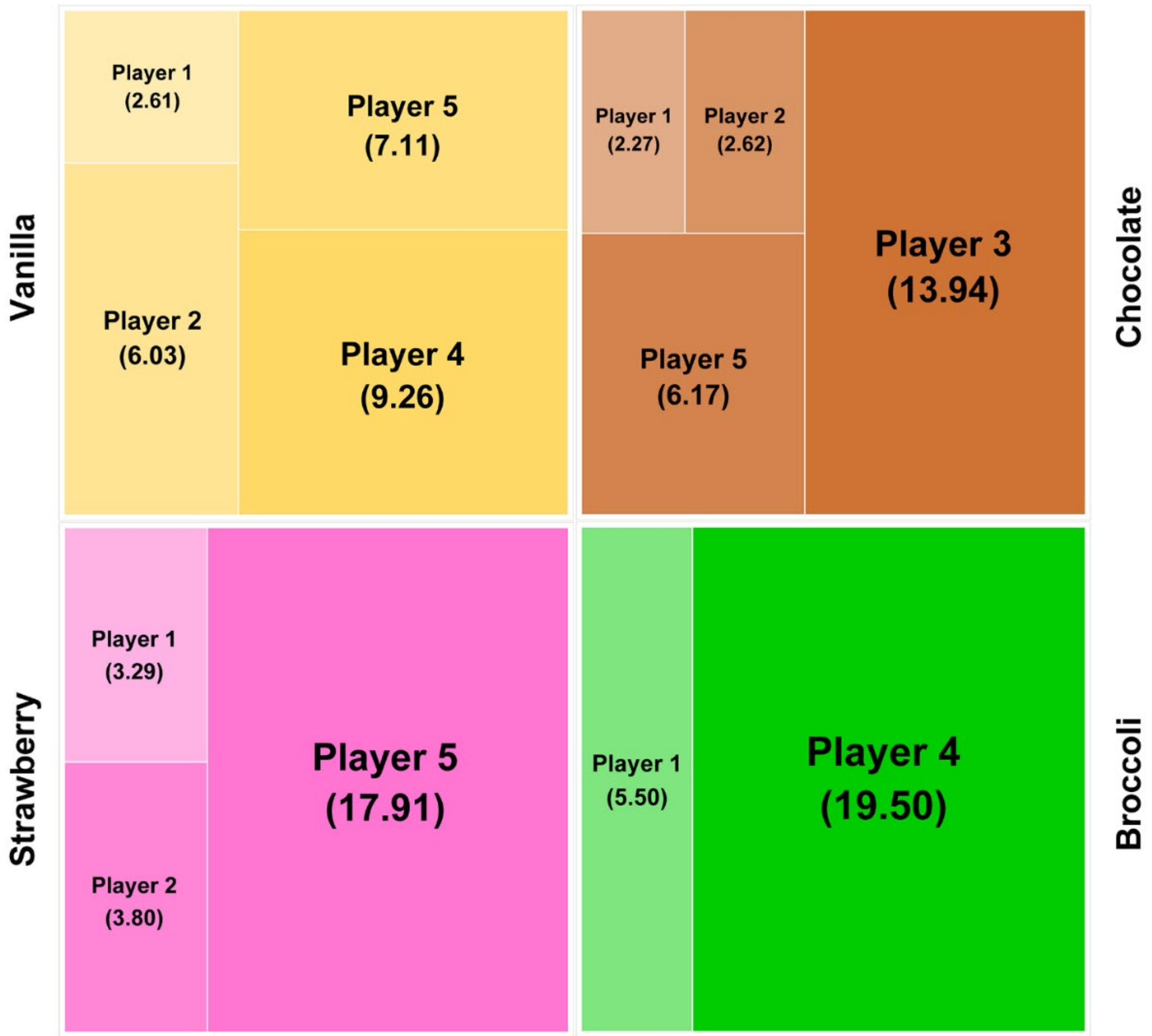

**Fig. 4 | Heterogeneous cake-cutting using the weighted Boltzmann distribution.** It is assumed that the whole cake (100%) has four flavors (vanilla, chocolate, strawberry, and broccoli) of equal size (25%). The vanilla and chocolate flavors are preferred by four players and thus shared by them. On the other hand, the least popular broccoli flavor is preferred by only two players; therefore, relatively large shares of the broccoli flavors are allocated to the two players.

**Table 1 | Homogeneous cake-cutting using the Boltzmann distribution and proportionality (Egalitarian, Prop I (Capitalist), Prop II (Socialist)).**

| Player | $E_j$ | $D_j$ | Boltzmann ($N_j$) | Def | Utility | Egalitarian (N) | Def | Utility | Prop I ($N_j$)* | Def | Utility | Prop II ($N_j$)** | Def | Utility |
|---|---|---|---|---|---|---|---|---|---|---|---|---|---|---|
| **1** | 5 | 4 | 12.17 | 8.17 | 0.9955 | 20 | 16 | 0.9999 | 5 | 1 | 0.8483 | 3.20 | -0.80 | 0.6640 |
| **2** | 10 | 10 | 14.06 | 4.06 | 0.8866 | 20 | 10 | 0.9640 | 10 | 0 | 0.7616 | 8.00 | -2.00 | 0.6640 |
| **3** | 20 | 24 | 18.75 | -5.25 | 0.6535 | 20 | -4 | 0.6823 | 20 | -4 | 0.6823 | 19.20 | -4.80 | 0.6640 |
| **4** | 25 | 34 | 21.66 | -12.34 | 0.5628 | 20 | -14 | 0.5286 | 25 | -9 | 0.6263 | 27.20 | -6.80 | 0.6640 |
| **5** | 40 | 53 | 33.36 | -19.64 | 0.5576 | 20 | -33 | 0.3604 | 40 | -13 | 0.6380 | 42.40 | -10.60 | 0.6640 |
| **Total utility** | | | **3.66 / 5.00** | | | **3.54 / 5.00** | | | **3.56 / 5.00** | | | **3.32 / 5.00** | | |

$E_j$ = Contribution to cake production of player *j*

$D_j$ = Need for cake of player *j*

Def (Deficiency) = Allocated share of cake ($N_j$) – Need ($D_j$)

* Capitalist: players receive cake units linearly proportional to their contributions ($E_j$)

** Socialist: players receive cake units linearly proportional to their needs ($D_j$)

**Table 2 | Heterogeneous cake-cutting using the weighted Boltzmann distribution.**

| Player | $E_j$ | $D_j$ | Homogeneous ($N_j$) | Heterogeneous | | | | | Diff |
|---|---|---|---|---|---|---|---|---|---|
| | | | | Vanilla ($N_j^1$) | Chocolate ($N_j^2$) | Strawberry ($N_j^3$) | Broccoli ($N_j^4$) | Total ($N_j$) | |
| **1** | 5 | 4 | 12.17 | 2.61 | 2.27 | 3.29 | 5.50 | 13.67 | 1.50 |
| **2** | 10 | 10 | 14.06 | 6.03 | 2.62 | 3.80 | 0.00 | 12.44 | -1.62 |
| **3** | 20 | 24 | 18.75 | 0.00 | 13.94 | 0.00 | 0.00 | 13.94 | -4.81 |
| **4** | 25 | 34 | 21.66 | 9.26 | 0.00 | 0.00 | 19.50 | 28.75 | 7.10 |
| **5** | 40 | 53 | 33.36 | 7.11 | 6.17 | 17.91 | 0.00 | 31.19 | -2.17 |
| **Total** | 100 | 125 | 100 | 25 | 25 | 25 | 25 | 100 | 0 |

$E_j$ = Contribution to cake production of player *j*

$D_j$ = Need for cake of player *j*

$N_j^i$ = cake share of flavor *i* allocated to player *j*

Diff (Difference) = Heterogeneous ($N_j$) – Homogeneous ($N_j$)

**Supplementary Information for**

# The Boltzmann fair division for distributive justice

**Ji-Won Park[1,2]*, Jaeup U. Kim[3], Cheol-Min Ghim[3], Chae Un Kim[3]***

[1]Department of Regional Science, Cornell University, Ithaca, NY 14853, USA

[2]Department of Economics, University of Ulsan, Ulsan 44610, Korea

[3]Department of Physics, Ulsan National Institute of Science and Technology (UNIST), Ulsan 44919, Korea

*** Corresponding authors: Ji-Won Park and Chae Un Kim**.

Email: jp429@cornell.edu
Email: cukim@unist.ac.kr

**Existence of finite $\beta$ value maximizing the total utility function in the homogeneous Boltzmann fair division**

As long as the utility function for each player is upper-bounded, the total utility is a decreasing function at a large enough $\beta$ where the player with the highest contribution receives most of the cake units. This observation indicates that the total utility function is maximized at a nonzero $\beta$ if it is an increasing function near $\beta = 0$. To clarify this condition, let us first analyze the behavior of the utility function at a small $\beta$ regime. Using Taylor expansion, the Boltzmann probability for the homogenous cake-cutting (Eq. 1 in Methods) can be approximated as

$$P_j = \frac{1+\beta E_j}{n+\sum_{k=1}^{n}\beta E_k} + O\left(\beta^2\right), \text{ for } j = 1,2,\cdots,n \tag{S1}$$

Then, the number of cake units allocated to each player is further approximated as

$$N_j = \frac{\tilde{N}}{n}\left(1+\beta E_j - \beta \overline{E}\right) + O\left(\beta^2\right), \text{ for } j = 1,2,\cdots,n \tag{S2}$$

where $\overline{E} \equiv \sum_{k=1}^{n} E_k / n$.

Now, let us consider the behavior of a utility function $u_j\left(x\right)$ where $x$ is the number of cake units allocated to player $j$. Note that the utility function $u_j\left(x\right)$ here is a nonlinear function (not necessarily a hyperbolic tangent) that reflects the realistic utility of the participating players. At $\beta = 0$, each player receives the same amount of cake units, $x = \tilde{N}/n$, and thus it is helpful to apply the Taylor expansion near this uniform division $x$ value,

$$u_j\left(x\right) = A_j + C_j\left(x-\frac{\tilde{N}}{n}\right) + O\left(\left(x-\frac{\tilde{N}}{n}\right)^2\right) \tag{S3}$$

where $A_j$ and $C_j$ are the value and slope of $u_j(x)$ at $x = \tilde{N}/n$, respectively.

Using Eq. S2, the utility function of player *j* and the total utility function (*U*) up to the first order in $\beta$ is given by

$$u_j(N_j) \cong A_j + \frac{\tilde{N}\beta}{n}\left(C_j E_j - C_j \overline{E}\right), \quad U = \sum_{j=1}^{n} u_j(N_j) \cong U_0 + \tilde{N}\beta\left(\frac{1}{n}\sum_{j=1}^{n} C_j E_j - \overline{C}\overline{E}\right) \quad \text{(S4)}$$

where $U_0 \equiv \sum_{j=1}^{n} A_j$ and $\overline{C} \equiv \sum_{j=1}^{n} C_j / n$.

This result indicates that the total utility (*U*) becomes an increasing function of $\beta$ and that its maximization at nonzero $\beta$ is guaranteed when the following condition is satisfied.

$$\frac{1}{n}\sum_{j=1}^{n} C_j E_j > \overline{C}\overline{E} \quad \text{(S5)}$$

Such a condition (Eq. S5) is not always satisfied for an arbitrary form of utility function. In this paper, we used a hyperbolic tangent-shaped utility function whose saturation maximum value is set at 1 for each player (Fig S1). In this case, if the cake is large enough relative to the players' needs so that the individual utility function is nearly saturated for some players, the condition (Eq. S5) can be satisfied when the players with higher contributions have higher needs. On the other hand, if the cake is small relative to the players' needs so that the individual utility is far from 1 for every player, the condition (Eq. S5) can be satisfied when the players with higher contributions have smaller needs.

For many realistic fair-division problems, the utility functions do not need to have the same saturation maximum for all players, and it might be possible that the heavy contributor's utility function has the tendency to have a higher slope ($C_j$) regardless of the number of allocated cake units. If the individual utility function of a heavy contributor has a higher slope and higher saturation maximum value (Fig. S2), then there will always exist a nonzero $\beta$ value maximizing the total utility.

Introducing heterogeneity to the preference of cake flavor drastically complicates the problem and each player's share is no longer equal to $\tilde{N} / n$ at $\beta = 0$. A simple analysis linearizing the utility function is not available, but the short conclusion mentioned above is still valid to a degree.

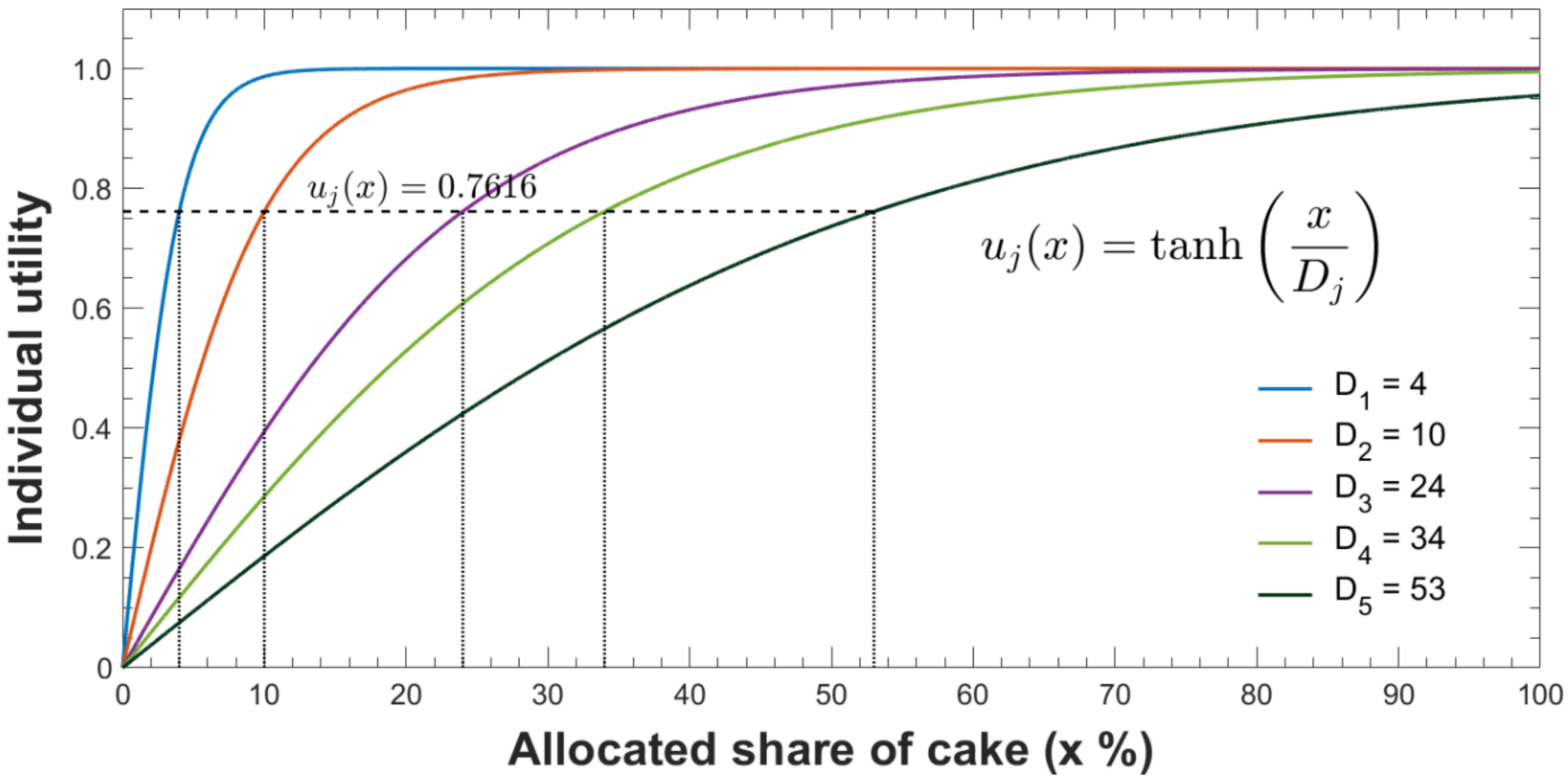

**Fig. S1.** Individual utility function for the five participating players. It was assumed that the players' needs are 4%, 10%, 24%, 34%, and 53%, respectively. Note that when the share of cake is the same as the players' needs, the utility function becomes saturated with the value, $u_j(D_j) = 0.7616$.

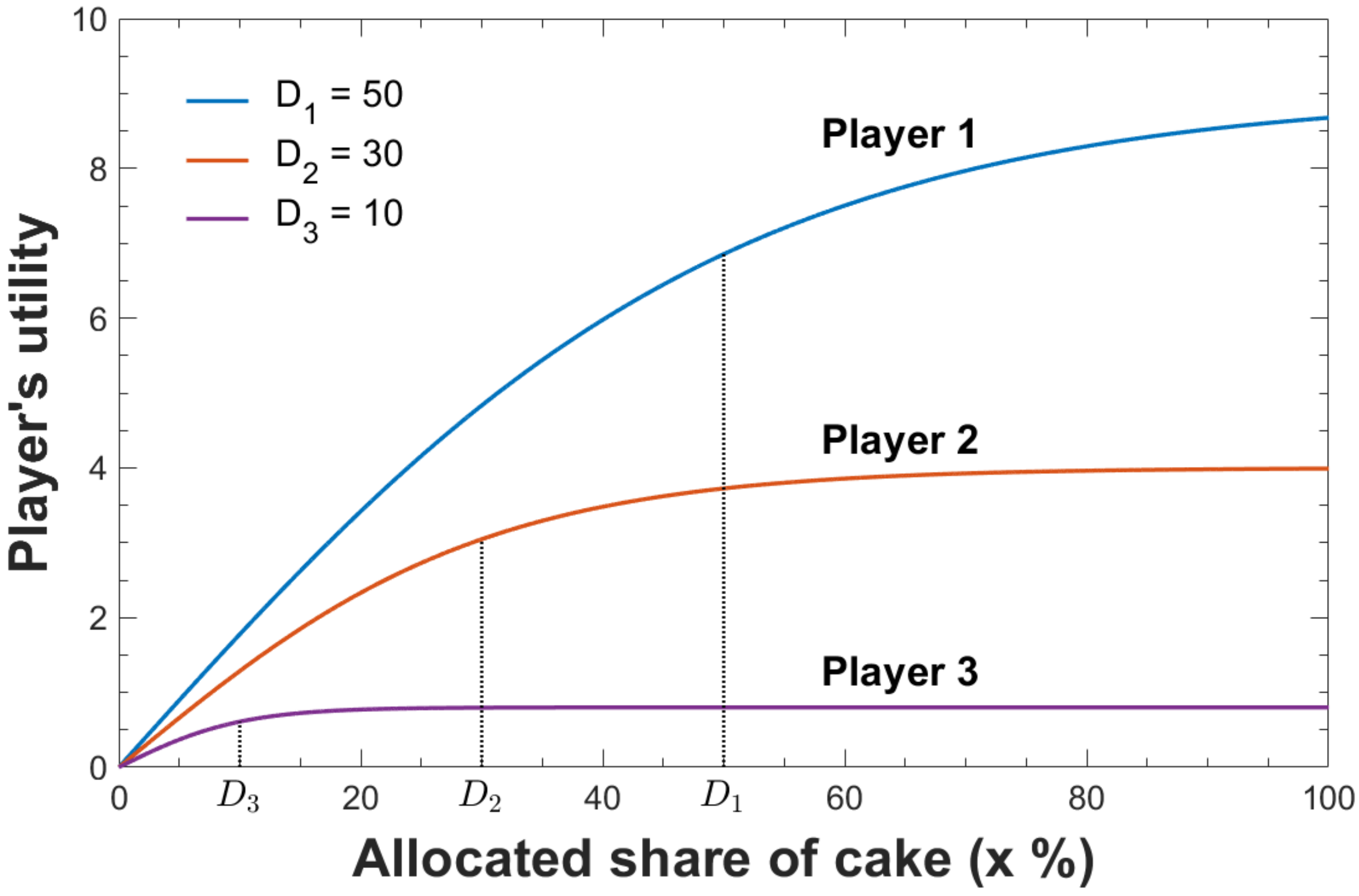

**Fig. S2.** Utility function for the three participating players. The saturation level of the hyperbolic tangent functions is set at 9, 4, and 0.8, respectively. As the division potential ($E_j$) reflecting player $j$'s contribution increases ($E_1 > E_2 > E_3$), the player's utility function has a higher slope at any allocated share of cake, thus leading to higher saturation values. In this case, the total utility ($U$) can always be maximized at a nonzero $\beta$ value. This situation would be suitable when the participating players represent a group of individuals, or communities and nations, that may have distinct population sizes and/or value systems.

**Table S1. Contributions, needs, and weight factors of five participating players.**

| Player | Contributions | Needs | Weight factor ( $w_j^i$ ) (in heterogeneous cake-cutting) | | | |
|---|---|---|---|---|---|---|
| | $E_j$ | $D_j$ | Vanilla | Chocolate | Strawberry | Broccoli |
| **1** | 5 | 4 | 0.25 | 0.25 | 0.25 | 0.25 |
| **2** | 10 | 10 | 0.5 | 0.25 | 0.25 | 0 |
| **3** | 20 | 24 | 0 | 1 | 0 | 0 |
| **4** | 25 | 34 | 0.5 | 0 | 0 | 0.5 |
| **5** | 40 | 53 | 0.25 | 0.25 | 0.5 | 0 |

Note that the constraint $\sum_{i=1}^{4} w_j^i = 1$ should be met for each player *j*.

**Table S2. Homogeneous cake-cutting using the Boltzmann distribution ($\beta^* = 0.0288$)**

| Player | $E_j$ | $D_j$ | $e^{\beta^* E_j}$ | $P_j$ | $N_j$ |
|---|---|---|---|---|---|
| **1** | 5 | 4 | 1.15 | 0.12 | 12.17 |
| **2** | 10 | 10 | 1.33 | 0.14 | 14.06 |
| **3** | 20 | 24 | 1.78 | 0.19 | 18.75 |
| **4** | 25 | 34 | 2.05 | 0.22 | 21.66 |
| **5** | 40 | 53 | 3.16 | 0.33 | 33.36 |
| **Sum** | **100** | **125** | **9.49** | **1.00** | **100** |

**Table S3. Heterogeneous cake-cutting using the Boltzmann distribution ( $\beta^*$ = 0.0286 )**

| Player | $w_j^i e^{\beta^* E_j}$ | | | | $P_j^i = \frac{w_j^i e^{\beta^* E_j}}{\sum_{k=1}^{n}\left(w_k^i e^{\beta^* E_k}\right)}$ | | | | $N_j^i$ | | | | $N_j$ |
|---|---|---|---|---|---|---|---|---|---|---|---|---|---|
| | **V** | **C** | **S** | **B** | **V** | **C** | **S** | **B** | **V** | **C** | **S** | **B** | **Total** |
| **1** | 0.29 | 0.29 | 0.29 | 0.29 | 0.10 | 0.09 | 0.13 | 0.22 | 2.61 | 2.27 | 3.29 | 5.50 | 13.67 |
| **2** | 0.67 | 0.33 | 0.33 | 0.00 | 0.24 | 0.10 | 0.15 | 0.00 | 6.03 | 2.62 | 3.80 | 0.00 | 12.44 |
| **3** | 0.00 | 1.77 | 0.00 | 0.00 | 0.00 | 0.56 | 0.00 | 0.00 | 0.00 | 13.94 | 0.00 | 0.00 | 13.94 |
| **4** | 1.02 | 0.00 | 0.00 | 1.02 | 0.37 | 0.00 | 0.00 | 0.78 | 9.26 | 0.00 | 0.00 | 19.50 | 28.75 |
| **5** | 0.78 | 0.78 | 1.57 | 0.00 | 0.28 | 0.25 | 0.72 | 0.00 | 7.11 | 6.17 | 17.91 | 0.00 | 31.19 |
| **Sum** | **2.76** | **3.18** | **2.19** | **1.31** | **1.00** | **1.00** | **1.00** | **1.00** | **25.00** | **25.00** | **25.00** | **25.00** | **100.00** |

V=Vanilla, C=Chocolate, S=Strawberry, B=Broccoli